\documentclass[]{aastex7}

\newcommand{\frb}{FRB\,20250316A}
\newcommand{\kms}{km~s$^{-1}$}

\newcommand{\beq}{\begin{equation}}
\newcommand{\eeq}{\end{equation}}

\newcommand{\cm}{cm$^{-2}$}
\newcommand{\Ms}{\textrm{M}_*}
\newcommand{\Msun}{\textrm{M}_\odot}
\newcommand{\kmps}{km~s$^{-1}$}
\newcommand{\hi}{H{\sc i}}
\newcommand{\MHI}{\rm{M_{HI}}}

\newcommand{\hii}{H{\sc i}\,21cm}

\begin{document}

\title{High-resolution Giant Metrewave Radio Telescope H{\sc i} 21\,cm imaging of the host galaxy of FRB\,20250316A}

\author{B. Kaur} 
\affiliation{Inter-University Centre for Astronomy and Astrophysics, Pune University, Pune 411007, India}
\email{balpreet.kaur@iucaa.in}

\author{N. Kanekar} 
\affiliation{National Centre for Radio Astrophysics, Tata Institute of Fundamental Research, Pune University, Pune 411007, India}
\email{nkanekar@ncra.tifr.res.in}

\author{J. X. Prochaska}
\affiliation{Department of Astronomy \& Astrophysics, UCO/Lick Observatory, University of California, 1156 High Street, Santa Cruz, CA 95064, USA}
\affiliation{Kavli Institute for the Physics and Mathematics of the Universe (Kavli IPMU), 5-1-5 Kashiwanoha, Kashiwa, 277-8583, Japan}
\email{xavier@ucolick.org}

\begin{abstract}
We report Giant Metrewave Radio Telescope (GMRT) H{\sc i}\,21cm imaging of NGC\,4141, the host galaxy of FRB\,20250316A at $z=0.0063$. Our GMRT H{\sc i}\,21cm images have spatial resolutions, at $z\approx0.0063$, of $\approx0.48-8.0$~kpc, and find evidence for (i)~a companion galaxy, LEDA\,2582852, to the south-west, (ii)~a nearby (27-kpc distant) H{\sc{i}} cloud to the south-west, (iii)~disturbances in the H{\sc{i}} distributions of both NGC\,4141 and LEDA\,2582852, and (iv)~high H{\sc{i}} column densities in the south-western outskirts of NGC\,4141. A Sloan Digital Sky Survey spectrum yields a low metallicity and a high star-formation rate (SFR) surface density in the south-western (SW) disk of NGC\,4141, and an  H$\alpha$-based SFR estimate that is significantly higher than that, at the same location, from the Galaxy Evolution Explorer near-ultraviolet image, indicating a recent burst of star-formation. The total SFR of NGC\,4141 is also found to be higher via the H$\alpha$ line than  from the 1.4~GHz radio continuum. The above evidence indicates that NGC\,4141 has recently (within the last $\approx 3$~Myr) acquired metal-poor gas, via either a merger or accretion, that resulted in the south-western starburst and that may have also triggered large-scale star-formation activity in NGC\,4141, resulting in the formation of the stellar progenitor of FRB\,20250316A and the other transients. Our highest-resolution (480~pc) GMRT H{\sc{i}} 21\,cm image finds no H{\sc{i}} 21\,cm emission from the location of FRB\,20250316A or the nearby star-forming region,  suggesting that most of the H{\sc{i}} here has been either ionized or converted into the molecular phase. Our non-detection of continuum emission at the location of FRB\,20250316A yields the $3\sigma$ upper limit $\nu L_{1.38~\rm GHz}<4.4\times10^{34}$~erg~s$^{-1}$, on the 1.38~GHz radio luminosity of a putative persistent radio source associated with FRB\,20250316A, one of the strongest constraints on the radio luminosity of such an associated persistent radio source. 
\end{abstract}


\section{Introduction} \label{sec:intro}

Fast Radio Bursts (FRBs) are millisecond-duration radio transient sources, with an isotropic sky distribution and high dispersion measures, indicating that they are extragalactic objects \citep[e.g.][]{Lorimer07,Chime2021,Law2024,Chime2025,Chime2025b,Shannon2025}. More than a thousand FRBs are known today, with tens of sources showing repeated bursts with similar properties \citep[e.g.][]{Spitler2016,Chime2019,Chime2023}. Radio interferometric studies have yielded accurate localizations and host galaxy identifications for $\approx 100$ FRBs out to $z \gtrsim 2$ \citep[e.g.][]{Chatterjee2017,Bannister2019, Ravi2019,Marcote2020,Ryder2023,Law2024,Caleb25}. The majority of 
these FRBs with high-probability associations 
arise in star-forming galaxies \citep[e.g.][]{Heintz2020,Gordon23frb,Sharma2024}, although a few have been detected in massive ellipticals and even a globular cluster \citep[e.g.][]{Kirsten2022,Sharma2023,Shah2025}.

Despite the large number of known FRBs, their origin remains a mystery today. The discovery of an FRB-like burst from a Galactic magnetar \citep{Bochanek2020,Chime2020b} implicates highly-magnetized neutron stars as one channel for FRB progenitors \citep[e.g.][]{Metzger2019,Margalit2019}. 
However, there remain a plethora of proposed emission mechanisms and sources which need ancillary constraints to reveal the predominant origins of FRBs. One such path is to study the local burst environment which describes the physical conditions that give rise to FRBs  and therefore offers insight into their origin. 

Unfortunately, the large distances of most FRBs make it difficult to examine their environments with high physical spatial resolution. Recently, \citet{frb25} reported the detection of the brightest FRB ever, FRB\,20250316A, with a very low dispersion measure, $\approx 161$~pc~cm$^{-3}$, indicating a nearby object. The CHIME/FRB Outrigger data yielded a localization with an accuracy of $\approx 68$~mas, coincident with NGC\,4141, a star-forming galaxy at $z \approx 0.0063$ \citep{frb25,Springob05}. The very low redshift of NGC~4141 and the accurate localization imply that the position of FRB\,20250316A is known to an accuracy of $\approx 11$~pc,\footnote{Throughout this paper, we will use a Tully-Fisher distance of 33.2~Mpc for NGC\,4141 \citep{Tully23}.} i.e. similar to the size of a molecular cloud. Here, we report high-resolution Giant Metrewave Radio Telescope (GMRT) \hii\ imaging of NGC\,4141, to study the neutral atomic gas conditions that gave rise to FRB\,20250316A.

\section{Observations and Data Analysis} 
\label{sec:obs}

\begin{table}
\centering
\begin{tabular}{|c|c|c|c|}
\hline
\hline
Angular Resolution 	 & Spatial Resolution & RMS noise        &  N$_{\rm HI}$~($3\sigma$) \\ 
($'' \times ''$)     & kpc~$\times$~kpc   & (mJy beam$^{-1}$) &  (cm$^{-2}$) \\
\hline
$50 \times 50$  	 & $8.0 \times 8.0$ & 0.65     & $8.9 \times 10^{18}$  \\
\hline
$30 \times 30$       & $4.8 \times 4.8$ & 0.49     & $1.9 \times 10^{19}$  \\
\hline
$15 \times 15$  	 & $2.4 \times 2.4$ & 0.45     & $ 6.9 \times 10^{19}$  \\
\hline
$10 \times 10$  	 & $1.6 \times 1.6$ & 0.38     &  $ 1.3 \times 10^{20}$\\
\hline
$6 \times 6$  	     & $0.95 \times 0.95$ & 0.30   &  $ 2.9 \times 10^{20}$ \\
\hline
$3 \times 3$  	     & $0.48 \times 0.48$ & 0.25   &  $9.6 \times 10^{20}$ \\
\hline
\hline
\end{tabular}
\caption{Parameters of the 6 \hii\ spectral cubes. The columns are (1)~angular resolution, in $'' \times ''$, (2)~the corresponding spatial resolution at $z = 0.0063$, in kpc~$\times$~kpc, (3)~the RMS noise per 10.4~\kms\ channel, in mJy~Beam$^{-1}$, and (4)~the $3\sigma$ \hi\ column density sensitivity at a velocity resolution of 10.4~\kms. The decrease in the RMS noise from a resolution of $50''$ to $3''$ is due to an increase in the number of baselines in the higher-resolution cubes. 
\label{table:cubes}}
\end{table}

We used the GMRT Band-5 receivers to image the \hii\ emission from NGC\,4141 on 2025 March~31, with 7 hours of on-source time (proposal DDTC429; PI: N. Kanekar). The GMRT Wideband Backend was used as the correlator, with a bandwidth of 200~MHz subdivided into 8,192 channels and covering the frequency range $1260-1460$~MHz. This yielded a velocity resolution of $\approx 5.2$~\kms\ at the redshifted \hii\ line frequency of $\approx 1411.51$~MHz. Observations of the compact source 1313+549, for 5m after every 40m on NGC\,4141, were used to measure the antenna-based complex gains and bandpass shapes, while the flux density scale was determined via observations of 3C286. 

The data were analysed in {\sc casa} \citep[version 6.6.5;][]{casa22}, following standard procedures. After initial data editing to remove non-working antennas and data affected by radio frequency interference (RFI), the antenna-based complex gains and bandpass shapes were measured using the routines {\sc gaincalR} and {\sc bandpassR} \citep{Chowdhury20,Chowdhury22b}. These gains were applied to the multi-channel target source visibilities, and the calibrated visibilities then split out. After further editing of data affected by  RFI, using a combination of manual flagging and the package {\sc aoflagger} \citep{Offringa12}, the visibility data were averaged to a resolution of $\approx 0.6$~MHz and then imaged using the task {\sc tclean}. The imaging used the w-projection and MT-MFS algorithms \citep{Cornwell08,Rau11}, to image a region of size $\approx 0.83^\circ$, well beyond the first null of the GMRT Band-5 primary beam. Following the imaging, we used a standard self-calibration procedure to refine the antenna-based complex gains, using 4 rounds of phase-only self-calibration and imaging, followed by 1 round of amplitude-and-phase self-calibration and imaging. The imaging during the self-calibration procedure used Briggs weighting, with robust parameter of -0.5, to obtain a narrower synthesized beam ($\approx 1\farcs7 \times 1\farcs6$). We then subtracted out the continuum image from the calibrated visibilities, using the task {\sc uvsub}, and carried out one more round of data editing with {\sc aoflagger}, based on the residual visibilities, before repeating the amplitude-and-phase self-calibration and making the final continuum image. This image, made with Briggs weighting and robust parameter of +1 (for higher sensitivity), has an angular resolution of $3\farcs2 \times 2\farcs9$ and an RMS noise of $\approx 8 \, \mu$Jy/Beam near the image centre. No continuum emission is detected from the FRB location; the $3\sigma$ upper limit on the 1.38~GHz FRB continuum flux density is $\approx 24 \, \mu$Jy. 

We next applied the antenna-based gains to the spectral-line visibilities, and then subtracted out the final continuum image from the calibrated visibilities, again using the task {\sc uvsub}. A final round of {\sc aoflagger} was then run on the residual spectral-line visibilities, with lower thresholds to remove weak RFI. The continuum model was then added back to the residual visibilities, to produce the final calibrated spectral-line dataset.

The GMRT contains baselines of length $\approx 0.07 - 25$~km, and is thus sensitive to \hii\ emission from a wide range of angular scales. We produced spectral cubes at a range of angular resolutions, $\approx 3\farcs0 - 50\farcs0$, to capture both the fine structure and the large-scale features. For each angular resolution, we initially made a continuum image at the same resolution and subtracted this out from the calibrated spectral-line visibilities, using the routine {\sc uvsub}. We then Hanning-smoothed and resampled the residual visibilities, to obtain a spectral resolution of $\approx 10.4$~\kms. Finally, for each resolution, we imaged the residual visibilities with appropriate tapering of the long baselines and Briggs weighting (with robust values of either 0.0 or +0.5) to produce the \hii\ cube. All spectral cubes were made at a velocity resolution of $\approx 10.4$~\kms. The cubes were cleaned down to a threshold of $\approx 0.5\sigma$, where $\sigma$ is the per-channel RMS noise for each cube. The parameters of the different \hii\ spectral cubes are listed in Table~\ref{table:cubes}.

We fitted a single-component Gaussian model to measure the 1.38~GHz flux density of NGC\,4141 from the $15\farcs0$-resolution continuum image (to avoid resolving out the extended continuum). This yielded a 1.38~GHz flux density of $2.45 \pm 0.15$~mJy. Similar values were obtained from the continuum images at resolutions of $10\farcs0$ and $30\farcs0$.

We compared the positions of unresolved sources in our highest-resolution continuum image (resolution $\approx 1\farcs7 \times 1\farcs6$) with their positions in the Very Large Array FIRST 1.4~GHz image \citep{Becker95}, using single-Gaussian models to identify the source position. The source positions in the two images were found to be consistent within $\approx 0\farcs5$. This is a conservative estimate of the astrometric accuracy of the GMRT images and spectral cubes.

Finally, we used the smooth and clip algorithm \citep{Serra12} in the source-finder software SoFiA \citep{Serra15,Westmeier21} to identify \hii\ emission features in the spectral cubes and generate the \hii\ moment images.
This algorithm compares the number of positive and negative features in each spectral cube at a range of spatial and spectral resolutions, to identify ``reliable'' positive emission features. For each cube, our search used Gaussian spatial kernels of 0, 3, and 6 pixels (with the synthesized beam of each cube sampled by 5 pixels), and boxcar spectral kernels of 0, 3, 5, 7, and 15 channels. We used a detection threshold of either $3.5\sigma$ or $4\sigma$ for the different cubes (chosen to obtain a sufficiently large number of  negative features, to ensure accurate reliability estimates), and a high reliability threshold of 0.9, to ensure that any identified emission features are likely to be real. 
We note that this is a conservative approach, and that there could be weak features in the cubes that are not identified at this reliability threshold. Conversely, this has the advantage that any identified features are very likely to be real. We note that, for the $30''$ resolution cube, we experimented with source-finding with SoFiA at lower reliability thresholds, down to a threshold of 0.7; no other candidate line emitters were detected in these searches within the FWHM of the GMRT primary beam. Finally, for all  emission regions identified by the SoFiA-based searches, we extracted \hii\ spectra from the cubes to confirm the reality of the features (see Fig.~\ref{fig:hi30_spc}).

\begin{figure}[t!]
\centering
\includegraphics[scale=0.2]{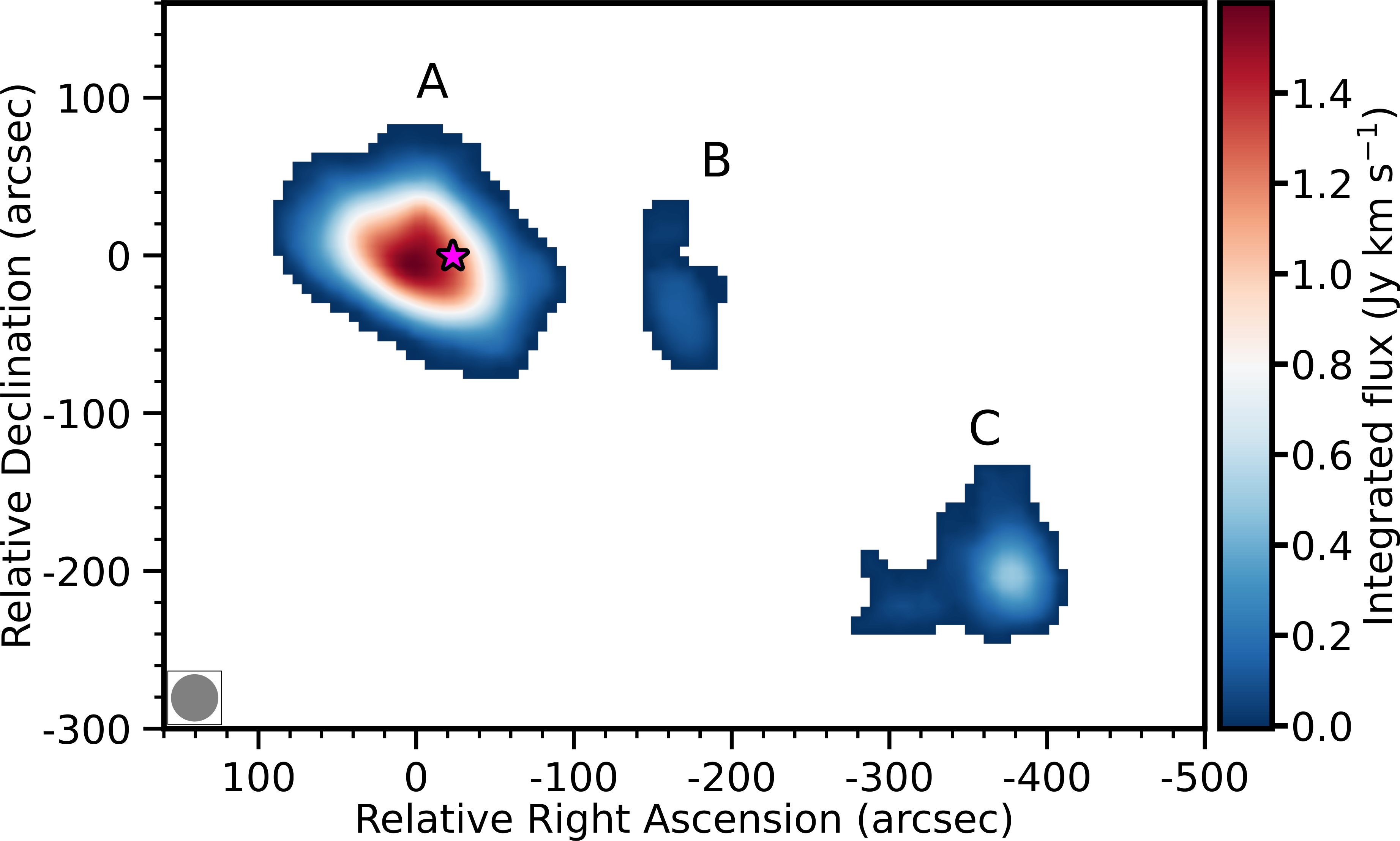}
\includegraphics[width=\textwidth]{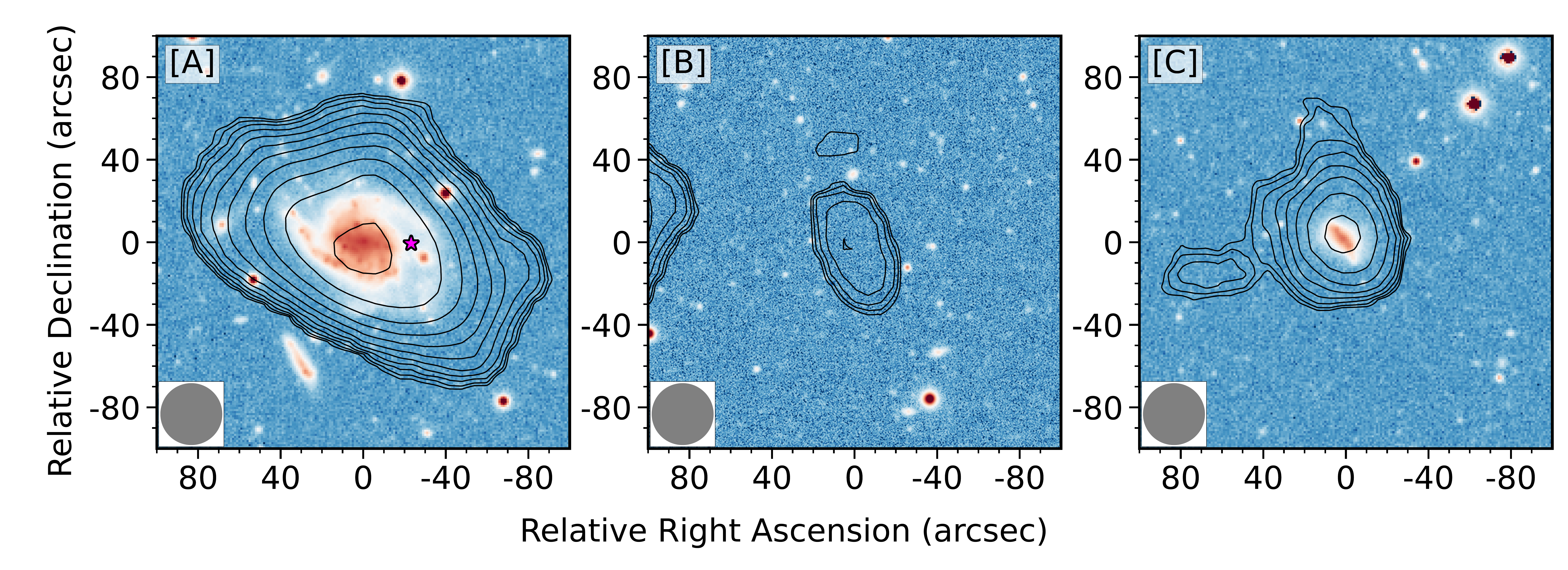}
\caption{Top panel: The GMRT wide-field velocity-integrated \hii\ moment-0 image at an angular resolution of $30\farcs0$, in colourscale. Three sources are clearly visible in the image, marked as A (NGC\,4141), B (a neighbouring \hi\ cloud), and C (LEDA\,2582852). The position of \frb\ is indicated by the magenta star. The bottom panels show the velocity-integrated \hii\ moment-0 images of the three objects (in contours), overlaid on the DECaLS g-band image \citep[colourscale;][]{Dey19}. In each bottom panel, the lowest contour is at the $5\sigma$ \hi\ column density sensitivity of the 30\farcs0-resolution spectral cube, $3.2 \times 10^{19}$~\cm\ per $10.4$~\kms\ channel, with subsequent contours increasing by factors of 1.5. In all panels, the grey circle in the inset at the bottom left indicates the GMRT synthesized beam.
\label{fig:hi30}}
\end{figure}

\begin{figure}[t!]
\centering
\includegraphics[width=\textwidth]{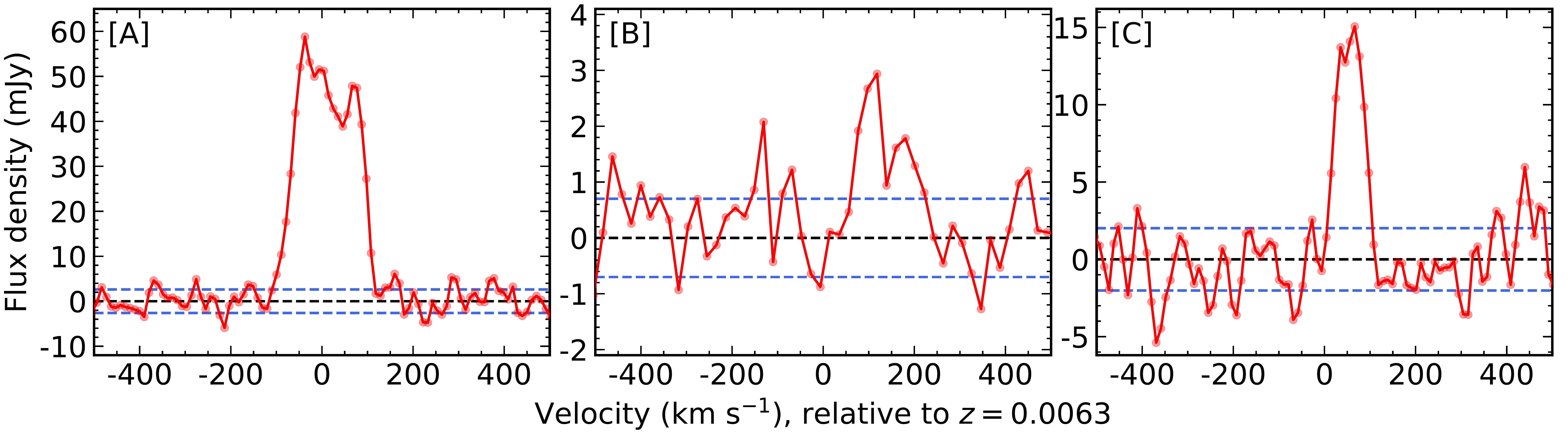}
\caption{The GMRT \hii\ spectra obtained from the 30\farcs0-resolution spectral cube, integrating over the spatial regions identified in Fig.~\ref{fig:hi30}. In each panel, flux density (in mJy) is plotted against barycentric velocity (in \kms), relative to $z = 0.0063$. Panels [A] and [C] show, respectively, the \hii\ spectra for NGC\,4141 and LEDA\,2582852, both at a velocity resolution of $\approx 10.4$~\kms, while Panel~[B] shows the \hii\ spectrum for the south-western \hi\ cloud, after Hanning-smoothing and resampling at a velocity resolution of $\approx 20.8$~\kms. The dashed blue lines indicate the $\pm1\sigma$ values for each spectrum, measured from the off-line regions.
\label{fig:hi30_spc}}
\end{figure}

\section{Results and Discussion} 
\label{sec:results}

\frb\ is located at R.A.=12h09m44.319s, Dec.=+58d50m56.708s, with an uncertainty of $\approx 68$~mas \citep{frb25}. This position lies within the disk of NGC\,4141, an SBc-type spiral galaxy at $z \approx 0.0063$. Besides \frb, a number of transients have been discovered in this galaxy over the last two decades, including two Type-II supernovae, 2009E and 2008X \citep[e.g.][]{Salo25}, an optical transient AT\,2025erx \citep{Andreoni25}, and an X-ray source recently localized with the Chandra X-ray Observatory \citep{Sun25a,Sun25b}. The number of detections of transient sources in NGC\,4141 suggests a recent increase in the star-formation activity in the galaxy.

\begin{figure}[b!]
\centering
\includegraphics[width=\textwidth]{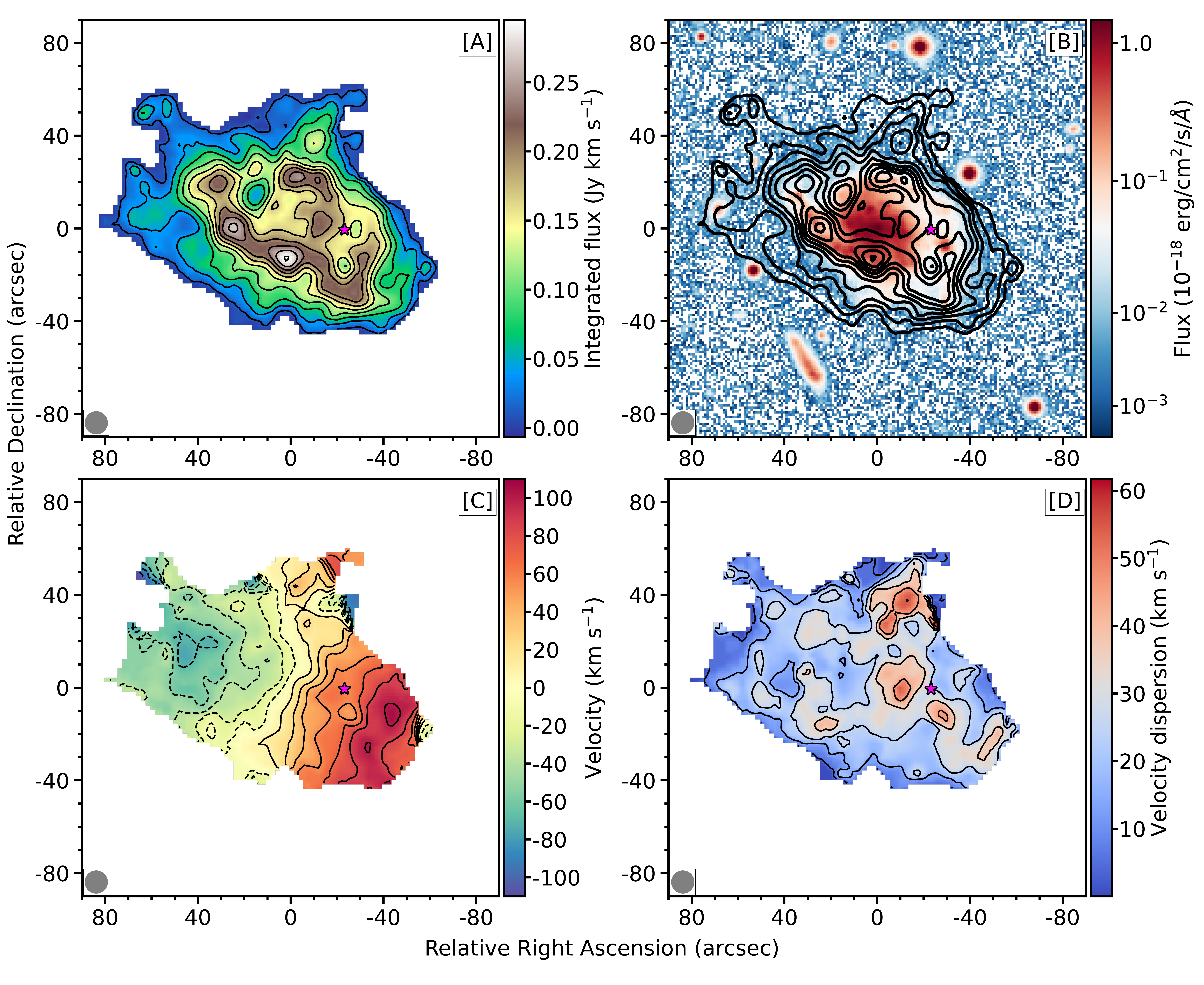}
\caption{GMRT \hii\ velocity moment images of NGC\,4141, at an angular resolution of $10\farcs0$. [A]~The velocity-integrated moment-0 image, in colourscale and contours. The lowest contour is at the $3\sigma$ \hi\ column density sensitivity of the spectral cube, $1.3 \times 10^{20}$~\cm\ per $10.4$~\kms, with subsequent contours increasing in steps of $3.5 \times 10^{20}$~\cm\ (i.e. $8\sigma$); the highest \hi\ column density contour is at  $3.2 \times 10^{21}$~\cm. [B]~The moment-0 image of [A] (contours), overlaid on the DECaLS g-band image \citep[colourscale; ][]{Dey19}. [C]~The \hii\ velocity field; the velocity contours are at intervals of 15~\kms, with negative velocities indicated by dashed contours. [D]~The \hii\ velocity dispersion map; the contours are at intervals of 10~\kms. In all panels, the axes are relative to the centre of NGC\,4141, with the magenta star indicating the location of \frb, and the grey circle in the inset at the bottom left indicating the GMRT synthesized beam.
\label{fig:hi10}}
\end{figure}

We used the GMRT 30\farcs0-resolution cube to carry out a wide-field search for \hii\ emission in and around NGC\,4141, as this provides a good balance between \hi\ column density sensitivity and image fidelity (note that the 50\farcs0-resolution image is made from a relatively small number of short GMRT baselines, which are more susceptible to systematic errors, e.g. due to RFI). The top panel of Figure~\ref{fig:hi30} shows the velocity-integrated \hii\ moment image at an angular resolution of 30\farcs0, obtained from our SoFiA smooth-and-clip search, with the coordinates relative to the location of NGC\,4141. Three distinct emission regions are visible here; the three lower panels of the figure show a zoom-in on each region, with the \hii\ emission shown in contours overlaid on the DECaLS g-band image \citep{Dey19}, in colourscale. \hii\ emission is detected from [A]~the FRB host galaxy, NGC\,4141,  [B]~an \hi\ cloud that is $\approx 2.8'$ (i.e. $\approx 27$~kpc) to the south-west of NGC\,4141, and [C]~a companion galaxy, LEDA\,2582852, $\approx 7.0'$ (i.e. $\approx 68$~kpc) to the south-west of NGC\,4141. The peak \hi\ column density of the south-western \hi\ cloud is $\approx 10^{20}$~\cm; this cloud does not have an optical counterpart in the DECaLS g-band image.  We note that all three emission regions are also clearly detected in the 50\farcs0-resolution cube.

\begin{figure}[t!]
\centering
\includegraphics[width=\textwidth]{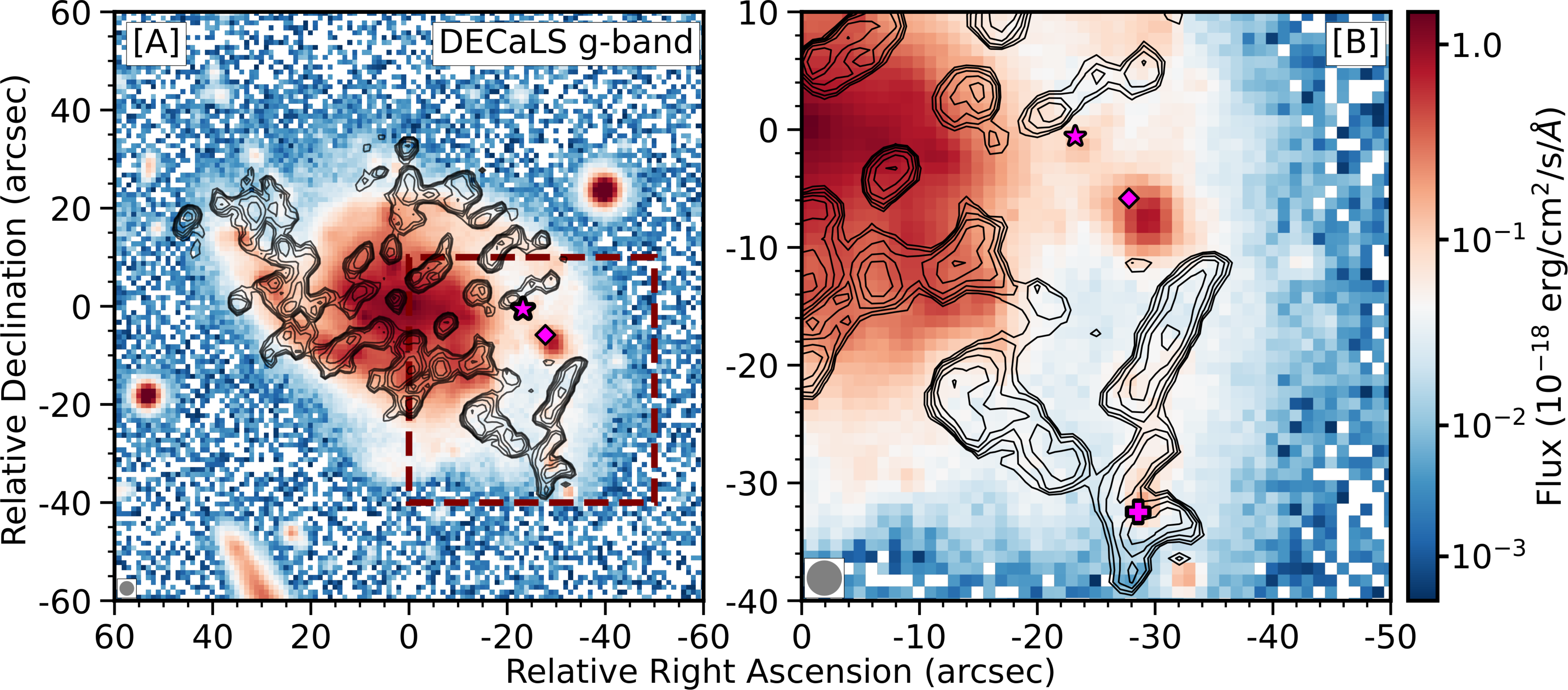}
\includegraphics[width=\textwidth]{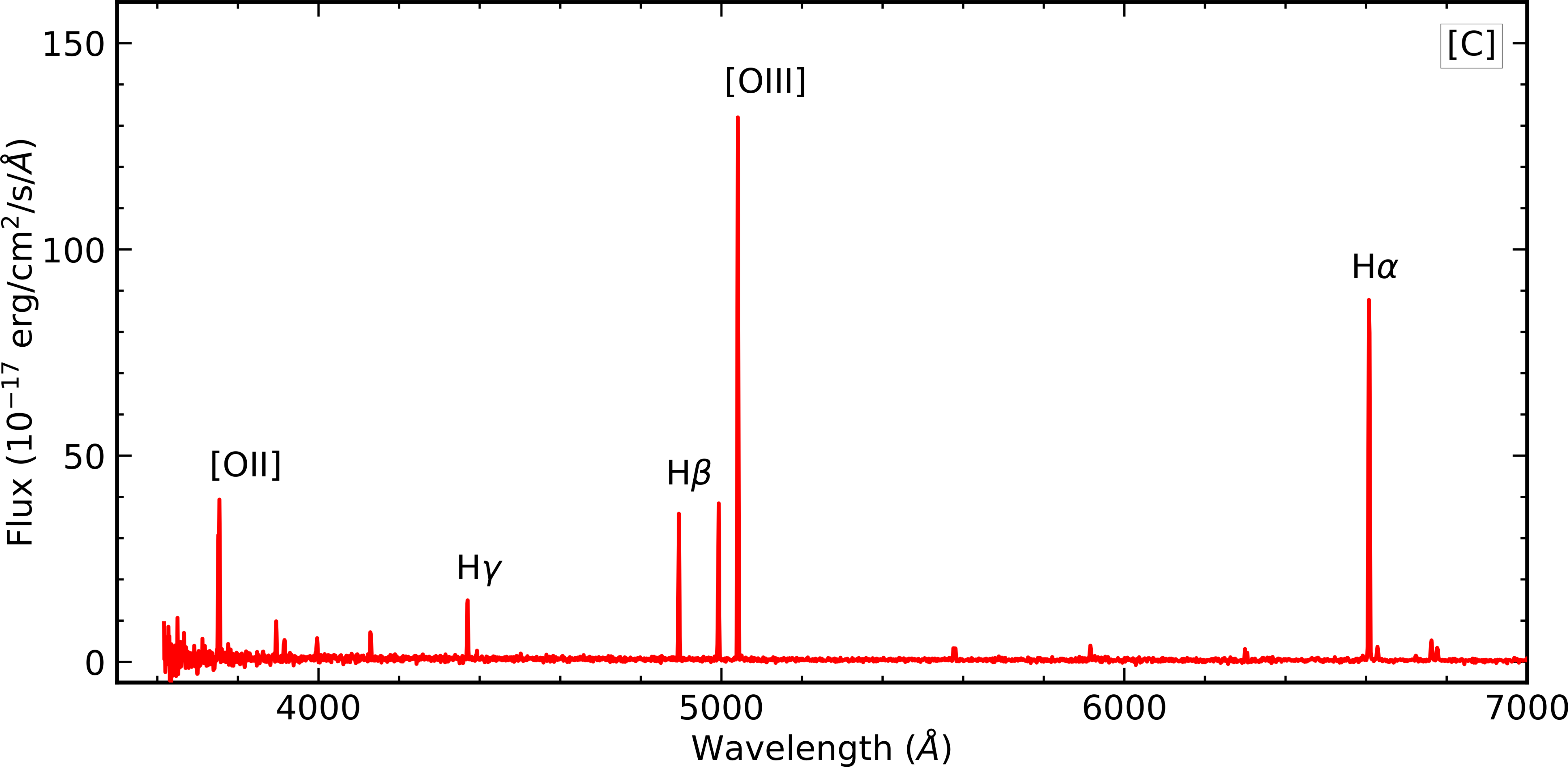}
\caption{The GMRT \hii\ moment-0 image at an angular resolution of $3\farcs0$ (contours), overlaid on the DECaLS g-band image \citep[colourscale; ][]{Dey19}. The axes are relative to the centre of NGC\,4141, with the magenta star and diamond indicating the locations of, respectively, \frb\ and the X-ray transient \citep{Sun25a}. The lowest contour is at the $3\sigma$ \hi\ column density sensitivity of the spectral cube, $9.6\times 10^{20}$~\cm\ per 10.4~\kms\ channel, with subsequent contours increasing by factors of 1.5, and the highest contour at an \hi\ column density of $7.3 \times 10^{21}$~\cm. [B]~A zoom-in on the maroon dashed square shown in [A]; the magenta plus sign indicates the location of the SDSS spectrum, shown in the bottom panel, [C].  In panels [A] and [B], the grey circle in the inset at the bottom left indicates the GMRT synthesized beam.
\label{fig:hi3}}
\end{figure}

\begin{figure}[t!]
\centering
\includegraphics[width=\textwidth]{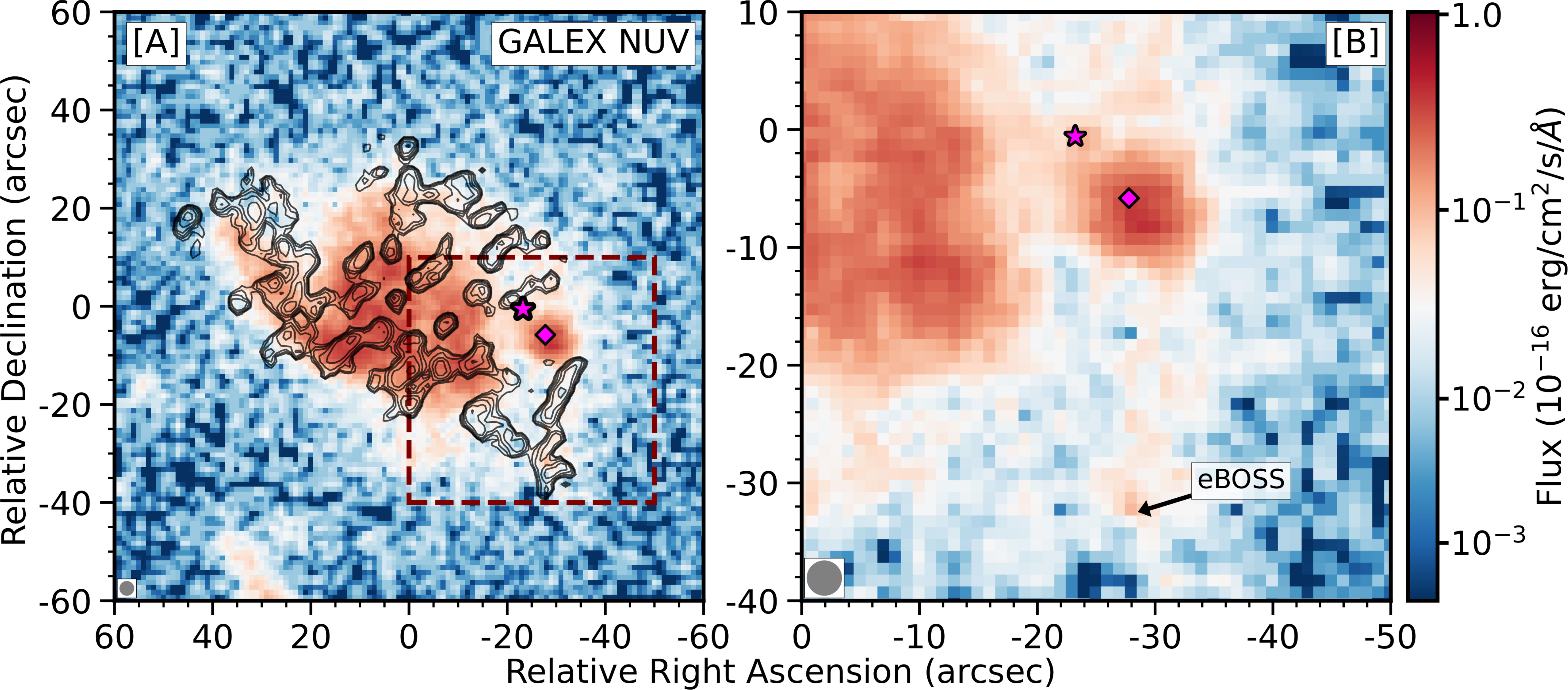}
\caption{[A]~The GALEX NUV image of NGC\,4141 \citep[colourscale; ][]{galex05}, overlaid on the GMRT  \hii\ moment-0 image at an angular resolution of $3\farcs0$  (contours). [B]~A zoom-in on the maroon dashed square shown in [A], without the \hi\ contours. The location of the eBOSS spectrum is indicated by the arrow. It is clear that only weak NUV emission is detected at this location. In panel [A], the grey circle in the inset at the bottom left indicates the GMRT synthesized beam.
\label{fig:galex}}
\end{figure}

The three panels of Figure~\ref{fig:hi30_spc} show the \hii\ spectra from [A]~NGC\,4141, [B]~the \hi\ cloud, and [C]~LEDA\,2582852, with \hii\ line flux density plotted against barycentric velocity, in \kms, relative to $z = 0.0063$. The \hii\ emission from both the \hi\ cloud and LEDA\,2582852 is seen to arise at velocities $\approx 50-100$~\kms\ redward of NGC\,4141. We obtain velocity-integrated \hii\ line flux densities of $8.56 \pm 0.14$~Jy~\kms\ (NGC\,4141), $0.307 \pm 0.038$~Jy~\kms\ (\hi\ cloud), and $1.386 \pm 0.091$~Jy~\kms\ (LEDA\,2582852) from the spatial regions identified in the SoFiA search, after correcting for the shape of the GMRT primary beam. Using the Tully-Fisher distance estimate of $33.2 \pm 0.5$~Mpc \citep{Tully23} for NGC\,4141, and assuming that the \hi\ cloud and LEDA\,2582852 are at the same distance, these line flux densities yield \hi\ masses of $(2.206 \pm 0.036) \times 10^9 \, \Msun$ (NGC\,4141), $(7.91 \pm 0.96) \times 10^7 \, \Msun$ (\hi\ cloud), and $(3.58 \pm 0.23) \times 10^8 \, \Msun$ (LEDA\,2582852). We note that the above uncertainties do not include systematic uncertainties in the flux density scale, which are typically $\approx 10$\% for the GMRT with the above calibration approach. Our measurement of the \hi\ mass of NGC\,4141 is in reasonable agreement with that of \citet{Springob05}, who obtain a velocity-integrated \hii\ line flux density of $7.79$~Jy~\kms\ [$\pm (10\%-15\%)$, from systematic uncertainties in the flux density scale] from an Effelsberg Telescope spectrum, after correcting for beam attenuation and pointing offsets, but not for possible self-absorption.

\begin{table}[t!]
\centering 
\caption{Properties of the three \hii-emitting regions identified via the SoFIA search of the 30\farcs0-resolution cube.
\label{table:frbprop}}
\begin{tabular}{lccc}
\hline
 & NGC 4141 & \hi\ cloud & LEDA\,2582852 \\
\hline
\hline
Right Ascension (J2000) &  12h09m47.32s & 12h09m25.86s & 12h08m58.73s     \\
Declination (J2000) & +58d50'57\farcs07 & +58d50'25\farcs7 & +58d47'31\farcs66 \\
Redshift & 0.0063  & 0.0066 & 0.0064                      \\
Tully-Fisher distance (Mpc) & $33.2 \pm 0.5$ & -- & -- \\
Separation (kpc) & -- & 27 & 68 \\
$\int S_{\rm HI} \, {\rm dV}$ (Jy~\kmps) & $8.56 \pm 0.14$ & $0.307 \pm 0.038$ & $1.386 \pm 0.091$ \\
$\MHI \, (\Msun)$ & $(2.206 \pm 0.036) \times 10^{9}$ & $(7.91 \pm 0.96) \times 10^7$ & $(3.58 \pm 0.23) \times 10^8$ \\
$\Ms \, (\Msun)$ & $1.6 \times 10^{9} \,^\star$ & -- & $3.8 \times 10^7\,^\dagger$   \\
SFR ($\Msun~\rm yr^{-1}$)& $0.57 \,^\star$  & -- & $0.016 \, ^\dagger$ \\
sSFR (yr$^{-1}$) & $3.6 \times 10^{-10}$  & -- & $4.3 \times 10^{-10}$ \\
$\MHI/\Ms \, $ & 1.4 & -- & 9.4 \\
$\tau_{\rm HI}$ (Gyr) & 3.9 & -- & 22.4 \\
\hline
\end{tabular}
\vskip 0.1in
References: $^\star$~\citet{Kewley05,Leroy2019}; $^\dagger$~\citet{Chang2015}. The quoted separations are 2-D ones, not taking into account the differences in the redshifts of the galaxies (as these could arise due to peculiar motions).
\end{table}

NGC\,4141 has a stellar mass of $1.6 \times 10^9 \, \Msun$ and a star-formation rate (SFR) of $\approx 0.57 \, \Msun$~year$^{-1}$ \citep[for a Chabrier initial mass function; ][]{Chabrier03,Kewley05,Leroy2019},  i.e. a specific star-formation rate (sSFR) of $\approx 3.6 \times 10^{-10}$~yr$^{-1}$. This implies an \hi-to-stellar mass ratio of $\approx 1.4$ and an \hi\ depletion timescale of $\approx 3.9$~Gyr. These are both within the spread of the corresponding values for galaxies of the xGASS sample with similar stellar masses \citep[e.g.][]{Saintonge17,Catinella2018}. Conversely, the companion galaxy, LEDA\,2582852 has an \hi-to-stellar mass ratio of $\approx 9.4$ and an \hi\ depletion timescale of $\approx 22.4$~Gyr; these indicate that LEDA\,2582852 is a relatively gas-rich galaxy. The \hi\ and stellar properties of NGC\,4141, the \hi\ clump, and LEDA\,2582852 are summarized in Table~\ref{table:frbprop}.

The four panels of Figure~\ref{fig:hi10} show the first three velocity moments of the \hii\ image of NGC\,4141, at an angular resolution of $10\farcs0$, produced by SoFiA searches using the smooth-and-clip algorithm \citep{Serra12,Serra15}.
Fig.~\ref{fig:hi10}[A] shows the moment-0 image (in contours and colourscale), while Fig.~\ref{fig:hi10}[B] shows the moment-0 image (in contours) overlaid on the DECaLS g-band image (in colourscale). In these figures, the highest \hi\ column densities are seen to arise from the central region of the galaxy, and the two prominent spiral arms. However, high \hi\ column densities are seen to also arise in the south-west of the galaxy, where not much stellar emission is seen in the DECaLS image. The \hi\ spatial distribution does not appear symmetric, with extended \hi\ wings clearly visible to the north, the east, and the north-east of the galaxy. An \hi\ hole, with column densities of $\lesssim 1.5 \times 10^{20}$~\cm\ can also be seen, just above the lower spiral arm and north-east of the galaxy centre. All of these signs of disturbances in the \hi\ spatial distribution suggest that the galaxy may have undergone a recent interaction. We note that the \hii\ moment-0 image of LEDA\,2582852 of Figure~\ref{fig:hi30}[C] shows extensions to the east and north, also suggesting a recent interaction. Conversely, the \hii\ velocity field of NGC\,4141, shown in Figure~\ref{fig:hi10}[C], appears broadly consistent with that expected from a rotating disk galaxy, albeit with some evidence of a turnover in velocity on the south-western side of the galaxy. Finally, the \hii\ velocity dispersion image of Fig.~\ref{fig:hi10}[D] shows that the highest velocity dispersions, of $\approx 45-60$~\kms, are seen around the northern \hi\ extension, around the galaxy centre (in the direction of \frb), and around the X-ray transient of \citet[][slightly south of \frb]{Sun25a}. These velocity dispersions are significantly higher than those typically seen in disk galaxies \citep[except for the central regions, e.g.][]{Walter08,deBlok24}. We note that the high observed velocity dispersions could arise due to beam-smearing, i.e. the crowding of iso-velocity contours within our angular resolution. Indeed, it is apparent from Figure~\ref{fig:hi10}[C] that the regions that show high velocity dispersions also show crowding of the iso-velocity contours. However, as discussed in detail by \citet{deBlok24} for the MHONGOOSE survey (for spatial resolutions similar to ours), such crowding of iso-velocity contours is not expected to be significant away from the central regions of galaxies. The observed high velocity dispersions and the crowding of iso-velocity contours in the outer regions of NGC\,4141 suggest that the galaxy is likely to have undergone a recent interaction.

Figure~\ref{fig:hi3}[A] shows the GMRT velocity-integrated \hii\ moment-0 image (contours) at a resolution of 3\farcs0, overlaid on the DECaLS g-band image \citep[colourscale; ][]{Dey19}. The top right panel, [B], shows a zoom-in on the region indicated by the dashed maroon square in [A]. While high column density gas can be seen north of, and close to, the location of \frb, no \hii\ emission is detected at the FRB location (see panel~[B]). We also extracted a spectrum at the location of \frb\ from the 3\farcs0-resolution cube, at a velocity resolution of $\approx 10.4$~\kms; the spectrum (not shown here) shows no evidence for \hii\ emission. Our non-detection of \hii\ emission at the FRB location yields a $3\sigma$ upper limit on the local \hi\ column density of $\approx 9 \times 10^{20}$~\cm, at a spatial resolution of $\approx 480$~pc. We note that \hii\ emission is also not detected from the region immediately south of \frb, where \citet{frb25} identify a star-forming clump; it appears that the bulk of the \hi\ in this region has been either ionized or converted to molecular hydrogen.

\begin{table}[t!]
\centering 
\caption{Measured flux densities and errors of a few selected lines detected in the SDSS eBOSS spectrum of Fig.~\ref{fig:hi3}[C]. Note that the quoted errors do not include uncertainties in the flux density scale.
\label{table:eboss}}
\begin{tabular}{lc}
\hline
\hline
Line &  Flux \\
     &  $\times 10^{-17}$~erg~s$^{-1}$~cm$^{-2}$   \\
\hline
[O{\sc ii}]$\lambda$3727\AA\  &  $171.8 \pm 2.2$ \\

H$\gamma$                     &  $42.02 \pm 0.62$ \\

[O{\sc iii}]$\lambda$4363\AA\ &  $5.63 \pm 0.62$ \\

H$\beta$                      &  $101.67 \pm 0.65$ \\

[O{\sc iii}]$\lambda$4959\AA\ &  $110.10 \pm 0.74$ \\

[O{\sc iii}]$\lambda$5007\AA\ &  $395.12 \pm 0.72$ \\

H$\alpha$                     &  $367.81 \pm 0.65$ \\

[N{\sc ii}]$\lambda$6584\AA\  &  $13.29 \pm 0.66$ \\
\hline
\hline
\end{tabular}
\end{table}

Remarkably, Figures~\ref{fig:hi3}[A] and [B] show that high \hi\ column density gas is also detected in the south-west of NGC\,4141, in the outer disk, which is quite faint in the DECaLS g-band image (see also Fig.~\ref{fig:hi10}[A] and [B]). Such high \hi\ column densities, $\approx 7.3 \times 10^{21}$~\cm, are typically found in the central, star-forming regions of galaxies, and not in the outskirts. We note that the high \hi\ column density gas in the south-western disk has a rough ``V''-shape, with the western arm of the ``V'' terminating close to the bright star-forming knot visible in the DECaLS g-band image (and the X-ray transient of \citealt{Sun25a}). 

Figure~\ref{fig:hi3}[C] shows a Sloan Digital Sky Survey (SDSS) extended Baryon Oscillation Spectroscopic Survey \citep[eBOSS; ][]{eboss16,eboss21} spectrum from the location indicated by the magenta plus symbol in Figure~\ref{fig:hi3}[B], close to the highest \hi\ column density in the south-western region of the galaxy. The eBOSS  spectrum shows strong [O{\sc iii}]$\lambda$5007\AA\ emission, and weaker H$\alpha$ and [O{\sc ii}]$\lambda$3727\AA\ emission. We emphasize that the central regions of NGC\,4141 and the region around \frb\ show stronger H$\alpha$ and [O{\sc ii}]$\lambda$3727\AA\ lines, and weaker [O{\sc iii}]$\lambda$5007\AA\ emission \citep[e.g.][]{frb25}, typical of a normal star-forming galaxy; it is only the south-western region that shows the strong [O{\sc iii}]$\lambda$5007\AA\ emission. Table~\ref{table:eboss} lists the observed flux densities of a number of the lines detected in the eBOSS spectrum.

To determine the SFR, the metallicity, etc., from the optical lines, one needs to first correct the line flux densities for dust extinction \citep[see ][for a similar analysis for the FRB environment]{frb25}. We obtain a Galactic E(B-V) value of $\approx 0.016$ for this direction from the IRSA dust extinction database\footnote{http://irsa.ipac.caltech.edu/applications/DUST} \citep{Schlafly11}, corresponding to a low visual extinction, $\rm A_V \approx 0.05$ for $\rm R_V = 3.1$. We hence ignore the Galactic extinction,  effectively simply including it in the extinction internal to NGC\,4141. We assume an intrinsic ratio of H$\alpha$ to H$\beta$ line luminosities of 2.87 \citep[e.g.][]{Osterbrock06}, and use the measured line flux densities to determine the visual extinction $\rm A_V$, assuming the G23 extinction model of \citet{Gordon23} \citep[see also][]{Gordon09,Gordon21,Fitzpatrick19,Decleir22} and $\rm R_V = 3.1$. This yields $\rm A_V = 0.66$. 

Applying the above $\rm A_V$ value and the G23 extinction model to our measured H$\alpha$ line flux density, we obtain a dust-corrected H$\alpha$ line luminosity of $L_{{\rm H}\alpha} = 7.83 \times 10^{38}$~erg~s$^{-1}$. Using the calibration of \citet{Kennicutt12}, this yields an SFR of $\approx 0.0042 \,\Msun$~yr$^{-1}$ at the location of the eBOSS spectrum, $\approx 7.0$~kpc from the centre of the galaxy, for a Chabrier initial mass function. The eBOSS fibres are circular, with diameter~$=2\farcs0$, implying an SFR surface density of $\approx 0.073 \, \Msun$~yr$^{-1}$~kpc$^{-2}$ for the region covered by the eBOSS fibre. This is an uncharacteristically high SFR surface density for the outskirts of a disk galaxy, and suggests that the recent burst of star-formation is likely to have been triggered by a merger or accretion event.

{Figure~\ref{fig:galex}[A] shows the Galaxy Evolution Explorer \citep[GALEX;][]{galex05} near-ultraviolet (NUV) image of NGC\,4141, in colourscale, with the velocity-integrated \hii\ emission from the $3\farcs0$-resolution cube overlaid in contours. Panel~[B] of the figure shows a zoom-in on the dashed maroon square shown in [A], without the \hii\ contours. While NUV emission is indeed detected around the location of the eBOSS spectrum (indicated by the arrow), this region does not stand out in any way. The measured NUV flux at the above location is $(3.3 \pm 1.2) \times 10^{-18}$~erg~s$^{-1}$~cm$^{-2}$~\AA$^{-1}$, after convolving the image to an angular resolution of $5\farcs6$, where the errors are from Poisson statistics. For the G23 extinction model and $\rm A_V=0.66$, this yields an SFR of $\approx 0.00038 \, M_\odot$~yr$^{-1}$. This is significantly lower than the SFR of $\approx 0.0042 \, M_\odot$~yr$^{-1}$ from the same location from the H$\alpha$ line. The apparent discrepancy can be explained by the fact that the mean age of the stellar population that contributes to the H$\alpha$ line emission is $\approx 3$~Myr, while that contributing to the NUV continuum emission is $\approx 10$~Myr \citep[e.g.][]{Kennicutt12}. Thus, a very recent episode of star-formation, on timescales $\ll 3$~Myr would yield a higher SFR from the H$\alpha$ line than the estimate from the NUV continuum. The lower SFR estimated from the GALEX NUV image of NGC\,4141 is thus consistent with a burst of star-formation, triggered by a recent merger or accretion event. In passing, we note that the angular resolution of the GALEX NUV image is a factor of $\approx 2.8$ larger than the $2\farcs0$ diameter of the eBOSS fibres. The SFR inferred from the NUV image is thus averaged over a larger region than that traced by the H$\alpha$ emission.

We next used the ``direct-T$_e$'' method to infer the metallicity at the location of the eBOSS spectrum \citep[e.g.][]{Izotov06,Jiang19}, using the extinction-corrected flux densities in the [O{\sc iii}]$\lambda$4363\AA, [O{\sc iii}]$\lambda$4959\AA, [O{\sc iii}]$\lambda$5007\AA, and [O{\sc ii}]$\lambda$3727\AA\ lines. This yields $12 + {\rm log(O/H)} = 7.87$. The metallicity at the galaxy centre is $12 + {\rm log(O/H)} = 8.74$ \citep{Kewley05}. This would imply a metallicity gradient of $\approx 0.13$~dex~kpc$^{-1}$ towards the south-west. Typical metallicity gradients in disk galaxies are $\approx 0.05$~dex~kpc$^{-1}$ \citep[e.g.][]{Zaritsky94}. A similar gradient, $\lesssim 0.05$~dex~kpc$^{-1}$, is obtained in NGC\,4141 from the slightly sub-solar metallicity measured by \citet{frb25} in the FRB environment. Therefore, the low metallicity at the location of the eBOSS spectrum instead suggests a recent addition of metal-poor gas, via either a merger or accretion from the circumgalactic medium. This fresh gas is likely to have resulted in the increased \hi\ column density seen in the south-western outskirts of the galaxy in Figs.~\ref{fig:hi3}[A] and [B], and to have triggered the recent star-formation activity, leading to the high inferred SFR surface density.

\citet{Leroy2019} find that NGC\,4141 lies $\approx 0.4$~dex above the local star-forming main sequence. Our estimate of the 1.38~GHz flux density ($2.45 \pm 0.15$~mJy) allows us to obtain an independent radio-based estimate of the SFR of NGC\,4141. Using the Tully-Fisher distance estimate of $33.2 \pm 0.5$~Mpc \citep{Tully23} for the galaxy, we obtain a radio-based SFR of $\approx 0.21 \, \Msun$~year$^{-1}$. This is lower than the SFR estimate of $\approx 0.57 \, \Msun$~year$^{-1}$ from the dust-corrected H$\alpha$ line luminosity \citep{Kewley05}. As noted by \citet{frb25}, this difference may arise due to a recent increase in the star-formation activity, as the radio SFR estimate is an average over 100~Myr, while the H$\alpha$-based estimate is an average over the last 10~Myr. The increased recent SFR may account for the detections of two Type-II SNe, an optical transient, an X-ray transient, and \frb, all within the last two decades, as well as for the fact that the galaxy lies above the local main sequence.

Overall, our GMRT \hii\ images find evidence for disturbances in the \hi\ distribution of NGC\,4141 and the companion, LEDA\,2582852, a nearby \hi\ cloud to the south-west without any discernible star-formation and at the velocities of the south-western side of the galaxy disk, and high \hi\ column densities in the south-western outskirts of the galaxy disk of NGC\,4141. The SDSS spectrum of the south-western region of the galaxy yields a high SFR surface density and a low metallicity, while the recent ($\lesssim 10$~Myr) SFR of NGC\,4141 is higher than the radio-based SFR averaged over 100 Myrs. Further, the SFR inferred from the GALEX NUV image at the location of the eBOSS spectrum is  significantly lower than that inferred from the H$\alpha$ line. All of these indicate that NGC\,4141 has recently acquired fresh gas, via either accretion or a merger, which compressed the \hi\ in the south-western part of the galaxy, triggering the starburst. This merger/accretion event may have also triggered additional star-formation activity in the disk of NGC\,4141, resulting in the increased recent SFR (from the comparison between the radio and the H$\alpha$ SFRs of the galaxy) and possibly in the formation of the stellar progenitor of \frb, and the progenitors of the other transients. Earlier \hii\ mapping studies of FRB host galaxies have also yielded evidence for recent merger activity in a number of systems \citep[e.g][]{Kaur22a,Glowacki24,Roxburgh25}, although we note that \citet{Roxburgh25} also found cases where the FRB host galaxy showed no clear evidence for a merger. It is possible that the merger (merger or accretion event, in the case of \frb) triggered a burst of star-formation activity that resulted in the formation of the massive stellar progenitors of the ensuing fast radio bursts. We emphasize, however, that the possibility that the progenitor of \frb\ was formed independent of the merger/accretion event cannot be ruled out.

Finally, our non-detection of continuum emission at the location of \frb\ places a $3\sigma$ upper limit of 24~$\mu$Jy on the 1.38~GHz flux density of any associated persistent radio source (PRS). This implies the $3\sigma$ upper limit of $\nu L_{1.38~\rm GHz} < 4.4 \times 10^{34}$~erg~s$^{-1}$ on the rest-frame 1.38~GHz radio luminosity of such a putative PRS. This is one of the strongest constraints on the radio luminosity of a PRS associated with a non-repeating FRB \citep[e.g.][]{Law24,frb25, An25}, and approximately four orders of magnitude lower than the radio luminosities of the four repeating FRBs that show a PRS \citep[e.g.][]{Chatterjee2017,Niu22,Bruni25,Zhang25}. For example (for our assumed Tully-Fisher distance of 33.2~Mpc to NGC~4141), \citet{An25} obtained $ \nu L_{15~\rm GHz} < 1.7 \times 10^{35}$~erg~s$^{-1}$, while \citet{frb25} obtained $ \nu L_{9.9~\rm GHz} < 1.4 \times 10^{35}$~erg~s$^{-1}$. As noted by \citet{An25}, the stringent constraints on the flux density of a putative PRS associated with \frb\ suggest, for pulsar wind nebula models \citep[e.g.][]{Dai2017}, an evolved progenitor in a low-density, weakly magnetized environment, rather than a young magnetar in a dense environment.

\section{Summary} \label{sec:sum}

We have used the GMRT Band-5 receivers to map \hii\ emission in and around NGC\,4141, the host of \frb, at $z = 0.0063$, at angular resolutions of $\approx 3\farcs0 - 50\farcs0$, i.e. spatial resolutions of $\approx 0.48 - 8.0$~kpc. The low redshift of NGC\,4141 implies that this is the finest spatial resolution at which \hii\ emission has ever been mapped from an FRB host galaxy. The GMRT \hii\ images find evidence for a companion galaxy, LEDA\,2582852, to the south-west of NGC\,4141, a nearby \hi\ cloud also towards the south-west,  high \hi\ column densities in the south-western outskirts of the galaxy disk, disturbances in the \hi\ distribution in both NGC\,4141 and LEDA\,2582852, and high velocity dispersions, $\gtrsim 50$~\kms, away from the central regions of NGC\,4141. All of these suggest a recent merger or accretion event that compressed the \hi\ in the south-western disk, and resulted in a burst of star-formation activity in the galaxy. Consistent with this, an SDSS spectrum in the south-west yields evidence of a low metallicity, suggesting recent addition of metal-poor gas, via either accretion or a merger. The estimate of the recent SFR in NGC\,4141, from the H$\alpha$ line, is also higher than our estimate of the radio-based SFR, averaged over 100-Myr timescales, while the SDSS spectrum finds a high SFR surface density in the south-western disk. Finally, the SFR estimate from the H$\alpha$ line in the south-western disk is significantly larger than the SFR estimate from the NUV continuum, indicating a very recent burst of star-formation, within the last $\approx 3$~Myr. We conclude that a merger or accretion event within the last $\approx 3$~Myr is likely to have compressed the \hi\ in the south-west of the galaxy and triggered the starburst activity. It is unclear whether the merger or accretion event was with the \hi\ cloud identified $\approx 27$~kpc away from NGC\,4141 or whether this cloud was ejected from NGC\,4141 by the merger/accretion event. Our highest-resolution \hii\ images, at a spatial resolution of $\approx 480$~pc at $z \approx 0.0063$, find no evidence of \hi\ at the location of \frb\ or in the neighbouring regions that show star-formation in recent KCWI images. While the location of \frb\ is away from the location of the south-western starburst,  the merger/accretion event is likely to have triggered star-formation activity across the galaxy (resulting in the higher galaxy-wide SFR estimate from the H$\alpha$ line compared to the radio-based SFR estimate), possibly causing the formation of the massive stellar progenitor of \frb. We obtain a $3\sigma$ upper limit of $\nu L_{1.38~\rm GHz} < 4.4 \times 10^{34}$~erg~s$^{-1}$ on the rest-frame 1.38~GHz luminosity of a persistent radio source associated with \frb, one of the strongest current constraints on the radio luminosity of a PRS associated with a non-repeating fast radio burst.

\begin{acknowledgements}
We thank an anonymous referee for comments that improved the manuscript. BK thanks Soumil Maulick, Rajeshwari Dutta and Anshuman Borghohain for discussions. We thank the staff of the GMRT who have made these observations possible. The GMRT is run by the National Centre for Radio Astrophysics of the Tata Institute of Fundamental Research. NK acknowledges support from the Department of Atomic Energy, under project 12-R\&D-TFR-5.02-0700, and from the Anusandhan National Research Foundation of the Department of Science and Technology, India, via a J. C. Bose Fellowship (JCB/2023/000030). 
J.X.P., as a member of the F$^4$ team, acknowledges support from NSF grant AST-1911140. 

Funding for the Sloan Digital Sky Survey IV has been provided by the Alfred P. Sloan Foundation, the U.S. Department of Energy Office of Science, and the Participating Institutions. 

SDSS-IV acknowledges support and resources from the Center for High Performance Computing at the University of Utah. The SDSS 
website is www.sdss4.org.

SDSS-IV is managed by the Astrophysical Research Consortium for the Participating Institutions of the SDSS Collaboration including 
the Brazilian Participation Group, the Carnegie Institution for Science, Carnegie Mellon University, Center for 
Astrophysics | Harvard \& Smithsonian, the Chilean Participation 
Group, the French Participation Group, Instituto de Astrof\'isica de Canarias, The Johns Hopkins University, Kavli Institute for the 
Physics and Mathematics of the Universe (IPMU) / University of 
Tokyo, the Korean Participation Group, Lawrence Berkeley National Laboratory, Leibniz Institut f\"ur Astrophysik Potsdam (AIP),  Max-Planck-Institut f\"ur Astronomie (MPIA Heidelberg), Max-Planck-Institut f\"ur Astrophysik (MPA Garching), Max-Planck-Institut f\"ur Extraterrestrische Physik (MPE), National Astronomical Observatories of China, New Mexico State University, New York University, University of Notre Dame, Observat\'ario 
Nacional / MCTI, The Ohio State University, Pennsylvania State 
University, Shanghai Astronomical Observatory, United 
Kingdom Participation Group, Universidad Nacional Aut\'onoma 
de M\'exico, University of Arizona, University of Colorado Boulder, University of Oxford, University of Portsmouth, University of Utah, University of Virginia, University 
of Washington, University of Wisconsin, Vanderbilt University, 
and Yale University.

The Legacy Surveys consist of three individual and complementary projects: the Dark Energy Camera Legacy Survey (DECaLS; Proposal ID \#2014B-0404; PIs: David Schlegel and Arjun Dey), the Beijing-Arizona Sky Survey (BASS; NOAO Prop. ID \#2015A-0801; PIs: Zhou Xu and Xiaohui Fan), and the Mayall z-band Legacy Survey (MzLS; Prop. ID \#2016A-0453; PI: Arjun Dey). DECaLS, BASS and MzLS together include data obtained, respectively, at the Blanco telescope, Cerro Tololo Inter-American Observatory, NSF’s NOIRLab; the Bok telescope, Steward Observatory, University of Arizona; and the Mayall telescope, Kitt Peak National Observatory, NOIRLab. Pipeline processing and analyses of the data were supported by NOIRLab and the Lawrence Berkeley National Laboratory (LBNL). The Legacy Surveys project is honored to be permitted to conduct astronomical research on Iolkam Du’ag (Kitt Peak), a mountain with particular significance to the Tohono O’odham Nation.

NOIRLab is operated by the Association of Universities for Research in Astronomy (AURA) under a cooperative agreement with the National Science Foundation. LBNL is managed by the Regents of the University of California under contract to the U.S. Department of Energy.

This project used data obtained with the Dark Energy Camera (DECam), which was constructed by the Dark Energy Survey (DES) collaboration. Funding for the DES Projects has been provided by the U.S. Department of Energy, the U.S. National Science Foundation, the Ministry of Science and Education of Spain, the Science and Technology Facilities Council of the United Kingdom, the Higher Education Funding Council for England, the National Center for Supercomputing Applications at the University of Illinois at Urbana-Champaign, the Kavli Institute of Cosmological Physics at the University of Chicago, Center for Cosmology and Astro-Particle Physics at the Ohio State University, the Mitchell Institute for Fundamental Physics and Astronomy at Texas A\&M University, Financiadora de Estudos e Projetos, Fundacao Carlos Chagas Filho de Amparo, Financiadora de Estudos e Projetos, Fundacao Carlos Chagas Filho de Amparo a Pesquisa do Estado do Rio de Janeiro, Conselho Nacional de Desenvolvimento Cientifico e Tecnologico and the Ministerio da Ciencia, Tecnologia e Inovacao, the Deutsche Forschungsgemeinschaft and the Collaborating Institutions in the Dark Energy Survey. The Collaborating Institutions are Argonne National Laboratory, the University of California at Santa Cruz, the University of Cambridge, Centro de Investigaciones Energeticas, Medioambientales y Tecnologicas-Madrid, the University of Chicago, University College London, the DES-Brazil Consortium, the University of Edinburgh, the Eidgenossische Technische Hochschule (ETH) Zurich, Fermi National Accelerator Laboratory, the University of Illinois at Urbana-Champaign, the Institut de Ciencies de l’Espai (IEEC/CSIC), the Institut de Fisica d’Altes Energies, Lawrence Berkeley National Laboratory, the Ludwig Maximilians Universitat Munchen and the associated Excellence Cluster Universe, the University of Michigan, NSF’s NOIRLab, the University of Nottingham, the Ohio State University, the University of Pennsylvania, the University of Portsmouth, SLAC National Accelerator Laboratory, Stanford University, the University of Sussex, and Texas A\&M University.

BASS is a key project of the Telescope Access Program (TAP), which has been funded by the National Astronomical Observatories of China, the Chinese Academy of Sciences (the Strategic Priority Research Program “The Emergence of Cosmological Structures” Grant \# XDB09000000), and the Special Fund for Astronomy from the Ministry of Finance. The BASS is also supported by the External Cooperation Program of Chinese Academy of Sciences (Grant \# 114A11KYSB20160057), and Chinese National Natural Science Foundation (Grant \# 12120101003, \# 11433005).

The Legacy Survey team makes use of data products from the Near-Earth Object Wide-field Infrared Survey Explorer (NEOWISE), which is a project of the Jet Propulsion Laboratory/California Institute of Technology. NEOWISE is funded by the National Aeronautics and Space Administration.

The Legacy Surveys imaging of the DESI footprint is supported by the Director, Office of Science, Office of High Energy Physics of the U.S. Department of Energy under Contract No. DE-AC02-05CH1123, by the National Energy Research Scientific Computing Center, a DOE Office of Science User Facility under the same contract; and by the U.S. National Science Foundation, Division of Astronomical Sciences under Contract No. AST-0950945 to NOAO.
\end{acknowledgements}

\facilities{GMRT}

\software{{\sc casa} \citep{casa22}; {\sc calR} \citep{Chowdhury21}; SoFiA \citep{Serra15}.}
\bibliography{bibliography}{}
\bibliographystyle{aasjournalv7}

\end{document}